# Atom motion in solids following nuclear transmutation

Gary S. Collins

Dept Physics and Astronomy, Washington State University, Pullman, WA, USA

collins@wsu.edu

**Abstract.** Following nuclear decay, a daughter atom in a solid will "stay in place" if the recoil energy is less than the threshold for displacement. At high temperature, it may subsequently undergo long-range diffusion or some other kind of atomic motion. In this paper, motion of $^{111}$Cd tracer probe atoms is reconsidered following electron-capture decay of $^{111}$In in the series of In$_3$R phases (R= rare-earth). The motion produces nuclear relaxation that was measured using the method of perturbed angular correlation. Previous measurements along the entire series of In$_3$R phases appeared to show a crossover between two diffusional regimes. While relaxation for R= Lu-Tb is consistent with a simple vacancy diffusion mechanism, relaxation for R= Nd-La is not. More recent measurements in Pd$_3$R phases demonstrate that the site-preference of the parent In-probe changes along the series and suggests that the same behavior occurs for daughter Cd-probes. The anomalous motion observed for R= Nd-La is attributed to "lanthanide expansion" occurring towards La end-member phases. For In$_3$La, the Cd-tracer is found to jump away from its original location on the In-sublattice in an extremely short time, of order 0.5 ns at 1000 K and 1.2 ms at room temperature, a residence time too short to be consistent with defect-mediated diffusion. Several scenarios that can explain the relaxation are presented based on the hypothesis that daughter Cd-probes first jump to neighboring interstitial sites and then are either trapped and immobilized, undergo long-range diffusion, or persist in a localized motion in a cage.

**Introduction**

Perturbed angular correlation of gamma rays (PAC) is a classic method for detecting nuclear hyperfine interactions, like Mössbauer effect and nuclear magnetic resonance [1]. In 2004, it was shown that PAC could be used to study atomic-motion of PAC probe atoms in solids when diffusive jumps lead to fluctuating electric-field gradients (EFG) [2]. This and subsequent related studies in this laboratory were carried out using the $^{111}$In radionuclide, that has a meanlife of 4.0 days and decays by electron-capture into an excited state of $^{111}$Cd, followed by emission of two gamma-rays in a cascade that has an intermediate 247 keV PAC level with long meanlife of 120 ns. PAC is used to measure the interaction of the local EFG with the quadrupole moment of the 247 keV level. The interaction is monitored in the time domain as a quadrupole perturbation function, or "PAC spectrum". For details about experimental methods, see [1] and [2] and references therein. Stochastic jumps of the probe atom produce fluctuating EFG's and damped perturbation functions.

The initial system studied was $^{111}$In in the highly-ordered intermetallic compound In$_3$La, that has the L1$_2$, or Cu$_3$Au, structure (Fig. 1) [2]. The In-sublattice includes the face-centered sites. Vacancies moving on that sublattice cause atoms to jump to neighboring lattice sites, as illustrated by arrows in the figure. Principal axes of the EFG tensors at the lattice sites are aligned along the X, Y, or Z cube axes, as shown, and it can be seen that each atomic jump to a near-neighbor site on the sublattice leads to reorientation of the EFG at the nucleus by 90°, causing decoherence in precessions of a nucleus when the mean residence time of the atom at a site is less than the lifetime of the PAC level. This is the simplest and most obvious vacancy diffusion mechanism for atoms on the In-sublattice.

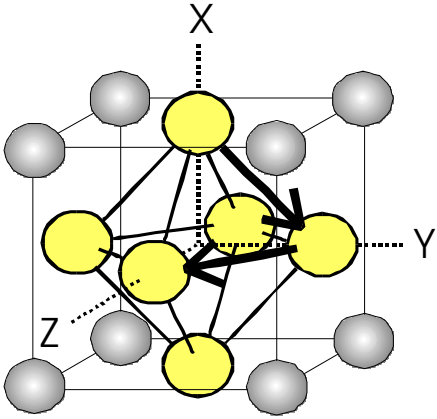

Fig. 1. L1$_2$ structure of In$_3$La. The In-sublattice has three orientationally inequivalent sites, with principal axes of the local EFG tensors oriented along the drawn X, Y and Z directions. Rapid passage of a vacancy on the In-sublattice causes reorientation of the EFG at the nucleus of a jumping atom by 90°. The fluctuating EFG leads to decoherence in precessions of the nuclear quadrupole moment of the PAC level of $^{111}$Cd. An alternative motion considered in this paper is of a Cd atom jumping from the In-sublattice to an interstitial lattice location such as at or near the octahedral interstitial site at the center of the unit cell.

In Fig. 2 are shown quadrupole perturbation functions for $^{111}$In/Cd in In$_3$La measured at different temperatures. Experimental spectra can be fitted with numerical spectra simulated for a specified
diffusion mechanism and used to obtain precise values of the mean jump-frequency (inverse of the mean residence time of probe atoms at a site). Fig. 2 shows experimental spectra measured at the indicated temperatures. Fig. 3 shows an Arrhenius plot of such fitted jump-frequencies measured for boundary compositions that were more In-rich ($x_1$) or In-poor ($x_2$). While In$_3$La and all the In$_3$R phases appear as "line compounds" in In-R binary phase diagrams, there is always a finite width to the phase field, of the order of a fraction of an atomic percent, so that, as can be seen from Fig. 3, even a small compositional difference can leads to a large difference in the jump-frequency [3]. By making measurements on slightly two-phase samples, e.g. with mean compositions of 24.5 and 25.5 at.% La, compositions of the very narrow L1$_2$ phase-field are stabilized against slight changes in the mean composition due to evaporation of one or other of the alloy constituents during day-long measurements [3]. For the In-rich data set in Fig. 3, the fitted jump-frequency activation enthalpy was 0.53(1) eV, with prefactor 1.02(14) THz [2]. Corresponding values for the In-poor set were 0.81 eV and 1.4(2) THz. The prefactors are of the order of typical vibrational frequencies of atoms in solids and are identified as jump-attempt frequencies.

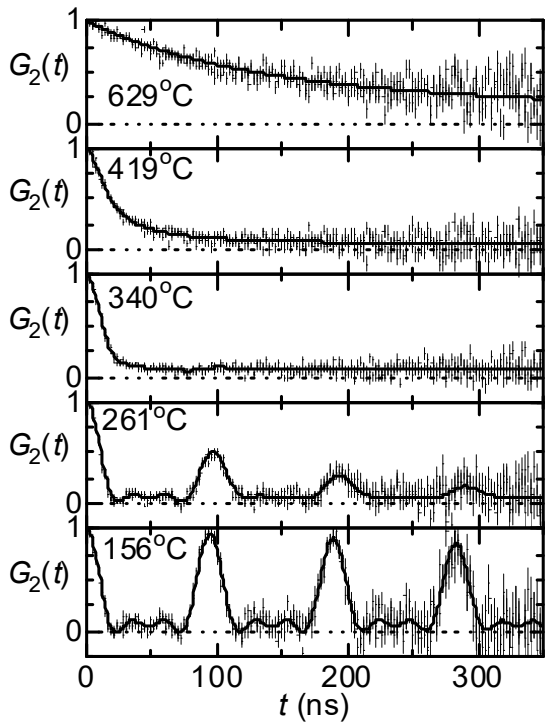

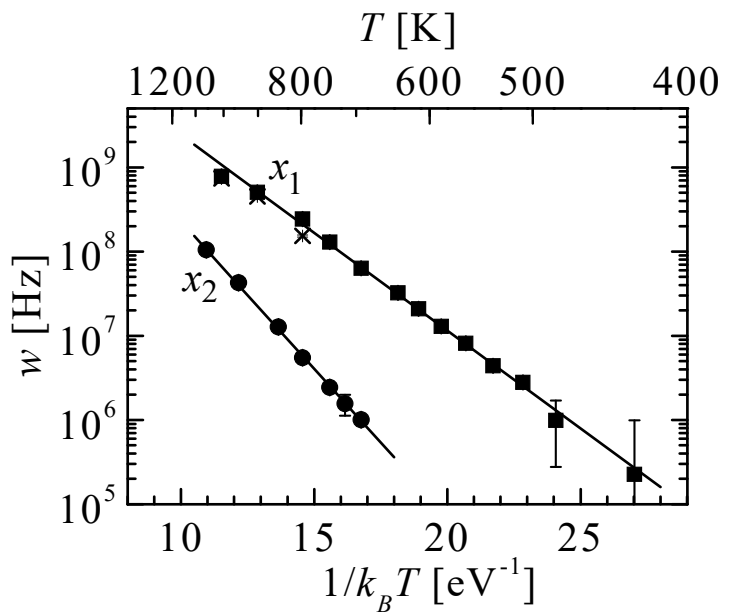

Fig. 2. PAC spectra for $^{111}$In/Cd probe atoms in In$_3$La having the In-rich boundary composition (from [2]). With increasing temperature, the nearly static perturbation function at 156°C becomes increasingly damped. Mean residence times of probe atoms on a site fitted assuming the simple vacancy diffusion model are, with increasing temperature, respectively, 2500, 130, 25, 10 and 1.3 ns. From [2].

Fig. 3. Arrhenius plots of relaxation-frequencies for In$_3$La samples (from [2]). Values are from fitting spectra in Fig. 2 assuming that the observed relaxation was caused by the simple vacancy diffusion mechanism. Data are shown for samples having In-rich ($x_1$) and In-poor ($x_2$) boundary compositions  It can be seen that jump-frequencies are 10-100 times greater for the In-rich sample. (Actual jump-frequencies are a factor of 2 greater than shown in the figure; see erratum in [2]).

The composition dependence of jump-frequencies shown in Fig. 3 is surprising. Concentrations of point-defects in compounds vary monotonically with composition so that, with increasing In-composition, the mole-fraction of In-vacancies ($V_{In}$) can only decrease. This trend rules out the simple In-sublattice vacancy diffusion mechanism as causing the observed nuclear relaxation in In$_3$La. Alternative mechanisms, such as a six-jump cycle involving a jump sequence between In- and La-sites [4] were originally suggested as possible explanations for the observations in ref. [2].

**Relaxation over the Series of In$_3$R Phases**

The initial study on In$_3$La was followed by systematic measurements over the entire series of rare-earth tri-indides [5], which all have L1$_2$ structure. Results are summarized as a function as a function of lattice parameter in Fig. 4 (from [5]). The vertical coordinate is the inverse of the temperature at which the jump-frequency is equal to 10 MHz, and was chosen as a convenient way to represent an entire Arrhenius plot of jump-frequencies with a single parameter rather than two (see [5] for details.) The jump-frequency is greater for In-rich boundary compositions (closed symbols) for R= La, Ce, Pr phases that have large lattice parameters, and greater for In-poor compositions (open symbols) for phases having smaller lattice parameters. This change in behavior is emphasized by dashed lines drawn on the figure that suggest trends extrapolated from either end of the series. Results shown in Fig. 4 were taken to indicate the existence of two distinct diffusion mechanisms, with crossover between the two mechanisms near the middle of the series, between Sm and Nd. Behavior for phases with smaller lattice parameters that have greater jump-frequencies in In-poor samples is consistent with the simple In-sublattice vacancy diffusion mechanism. This is because the In-vacancy mole fraction will be greater for In-poorer boundary compositions since point-defect mole fractions always vary monotonically with sample composition. Values of data shown in Fig. 4 are tabulated in the Appendix.

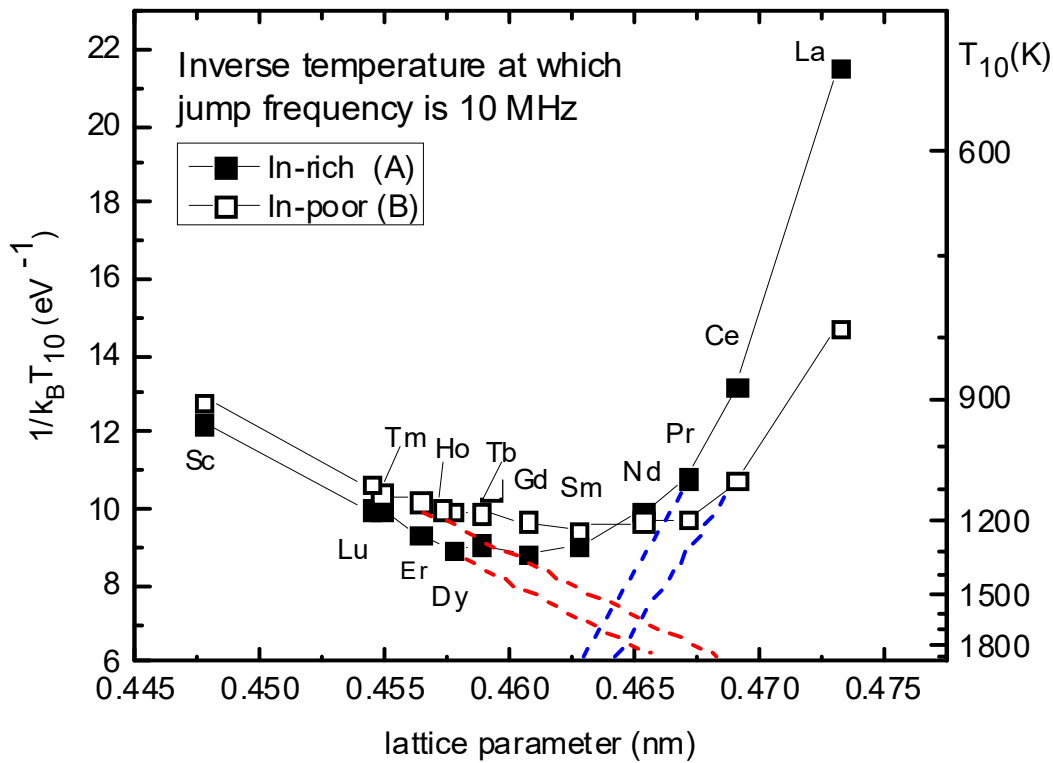

Fig. 4. Graph showing trends in jump-frequencies of $^{111}$Cd tracer atoms in In$_3$R phases (R= rare-earth), from [5]. A higher value of the y-coordinate indicates greater jump-frequencies. Data are plotted as a function of the lattice parameters of the phases. Dashed lines emphasize trends extrapolated from the two ends of the series.

Fig. 5 shows a conventional view of relaxation frequencies along the series at 800 K. The plot was obtained from the data shown in Fig. 4 and tabulated in the Appendix using the expression for thermally-activated relaxation,

$$w(T) = w_0 \exp(-Q/k_B T) \qquad (1)$$

in which $Q$ is the relaxation activation enthalpy and $w_0$ is a frequency prefactor, obtained in the usual way by fitting the temperature dependence of the relaxation frequency with eq. 1.

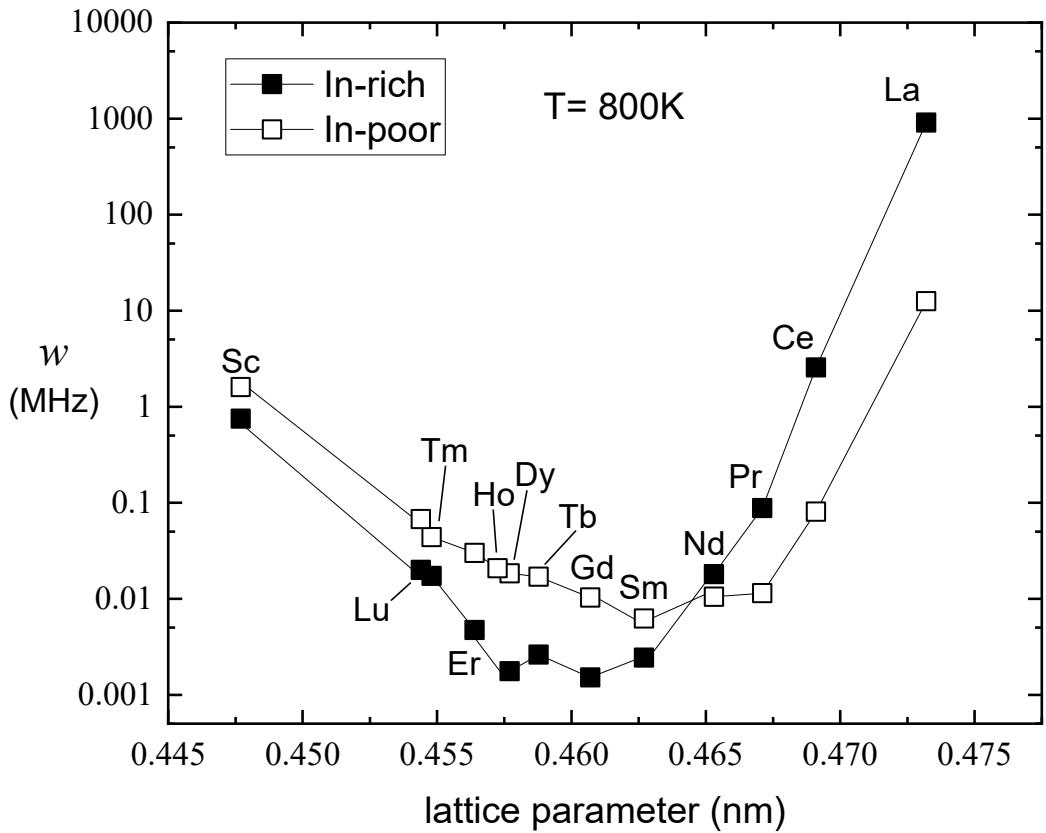

Fig. 5. Jump frequencies of $^{111}$Cd probe atoms at 800 K for opposing boundary compositions of L1$_2$ phases In$_3$R, in which R are the indicated rare-earth elements. Date were estimated as explained in the text from values shown in Fig. 4 and tabulated in the Appendix.

From many studies of L1$_2$ phases of rare-earth elements with In, Al, Ga and Sn, frequency prefactors of individual activation enthalpy data sets for all lattice parameters have been found to be in the range 1-100 THz [6]. This is taken to represent the vibrational frequency of the PAC probe atom, consistent for example with a vacancy diffusion mechanism. From eq. 1, one can then obtain the relaxation frequency in MHz at any temperature $T$ using values of $1/k_B T_{10}$ tabulated in the Appendix via the expression

$$w = w_0 \left( \frac{10 MHz}{w_0} \right)^{T_{10}/T} . \qquad (2)$$

This is illustrated in Fig. 5 for T= 800 K, calculated assuming the geometrical average value $w_0 \approx 10$ THz for all samples. As can be seen, frequency ratios for In-rich and In-poor compositions differ by factors of 3-10 for the phases having smaller lattice parameters. These factors suggest that the mole fractions of In-vacancies are 3-10 times greater for the In-poorer boundary compositions in those phases (see [5] for examples of complete Arrhenius plots). The frequency ratios are even greater for La, Ce and Pr phases as a consequence of a different relaxation mechanism discussed further below.

**Relaxation and Site-Preferences in the Series of Pd$_3$R Phases**

More recent measurements on the parallel series of Pd$_3$R phases of L1$_2$ structure [7] have given insight into the behavior for light lanthanide indides (R= La, Ce, Pr) [8]. The parent [111]In probes, which are impurities in Pd$_3$R phases, were found to strongly prefer to occupy Pd-sites in Pd-poor heavier lanthanide-palladides that have small lattice parameters (R= Lu...Tb), and to prefer R-sites in the Pd-richer light lanthanide-palladides that have larger lattice parameters (R= Eu...La). Indeed, [111]In probes were found to occupy only La-sites in either Pd-rich or Pd-poor Pd$_3$La. At the same time, nuclear relaxation (and consequently the jump-frequency) of Cd-daughter probes on Pd-sites was undetectably small at ~1000°C for the heavy palladides (R= Lu...Tb) while relaxation was appreciable and increased along the series R= Sm, Eu, Nd, Pr, Ce. This increase was taken to indicate increasing instability of Cd impurities on the Pd-sublattice following decay of In into Cd, caused by a change in the site-preference of Cd along the series [8]. The similarity in the increase in jump-frequencies in the light indides and palladides (see [8]) inferred that there are parallel changes in the site-preference of Cd-solutes in both the indide and palladide series with R= Nd, Pr, Ce, La. The mole-fraction of La-vacancies will be greater in In-rich samples, seemingly consistent with the jump-frequencies shown in Figs. 4 and 5. However, from Fig. 3, it can be seen that the jump-frequency for In-rich In$_3$La extrapolated to room temperature is 800 Hz [2], an extraordinarily high frequency for vacancy-mediated diffusion at low temperature since the vacancy jump-frequency itself would have to be much greater. Diffusion mechanisms consistent with a mechanism involving La-vacancies, such as the six-jump mechanism [4], are complex and have high activation enthalpies caused by formation of antisite atoms that appear to rule out large relaxation frequencies. It is concluded that vacancy-mediated diffusion mechanisms are unable to explain observed relaxation in the light-lanthanide indides (R= La, Ce, Pr).

For lack of alternatives, it is proposed that Cd-probes in light-lanthanide phases jump from the In- or Pd-sublattice to an interstitial lattice location, a scenario not considered in the seminal paper on In$_3$La [2] or in [5]. For further details, see [8]. Trends suggested in Fig. 4 by the dashed lines extrapolated from the domain of small lattice parameters indicate that the rate of jumps via the simple In-sublattice vacancy diffusion mechanism are negligible in comparison with the relaxation mechanism dominant for In$_3$R (R= La, Ce, Pr). Why the jump frequency due to the sublattice vacancy diffusion mechanism decreases so dramatically as the lattice parameter increases is unknown. Presumably it is a result of trends in vacancy formation and migration enthalpies along the series and is an interesting subject for future theoretical study. It should be noted that an alternative explanation of the observations has been proposed based on density-function theory calculations of the formation enthalpies of point defects in end-member phases In$_3$La and In$_3$Lu [9]. The interaction enthalpies between near-neighbor In- and R-vacancies were also calculated. An attractive association enthalpy of 0.2 eV was found for such vacancy pairs in In$_3$La that was absent in In$_3$Lu. This opens up the possibility that correlated divacancy migration might contribute to atomic motion in In$_3$La, but further computational work is needed.

## Lanthanide Expansion

Since the chemistries of rare-earth elements are so similar, differences in diffusion behavior along both $In_3R$ and $Pd_3R$ series are attributed primarily to variations in lattice parameters and accompanying size-effects. Fundamental identification of preferred lattice locations and diffusion mechanisms might be pursued using density-function theory methods such as Wien2k [13]. In the meantime, the idea that daughter Cd-atoms jump to interstitial sites in $In_3R$ and $Pd_3R$ phases are made plausible here using elementary considerations: (1) Lattice expansion along the lanthanide series going towards the La end-member phase is correlated directly with a volume increase of interstitial sites, and (2) the expanding lattice reduces steric hindrance surrounding In or Pd sites. In a manner of speaking, it is now shown that the lattice is "looser" for the La end-member phases. Consider atomic radii of host and probe atoms. These will be taken to equal half the distances of closest approach $d$ in the elemental metals; 3.25, 2.75, 3.73, 3.65, 3.43 and 2.98 angstroms, respectively, for In, Pd, La, Ce, Lu and Cd [10]. Since the rare-earth atoms are larger than Pd, In or Cd atoms, they control the spacings between atoms in the structure. Structural data for six phases are presented in Table 1. For a pure $L1_2$ phase $A_3B$, the distance along the close-packed <110> direction from atom B to B is equal to $\sqrt{2}a$ (length of a diagonal across a face of the unit cell in

Table 1. Structural parameters for selected phases. Lattice parameters $a$ and distances of nearest approach in elements are in angstroms, with lattice parameters from [11] and distances from [10].

| Phase $A_3B$ | Lattice par. $a$ | $d_A + d_B - \sqrt{2}a$ | $d_{In} + d_B - \sqrt{2}a$ | $d_A - a/\sqrt{2}$ | $\dfrac{d_A + d_{Cd}}{2} - a/\sqrt{2}$ |
|---|---|---|---|---|---|
| $In_3La$ | 4.7321 | 0.288 | (0.288) | -0.096 | -0.231 |
| $In_3Ce$ | 4.6876 | 0.271 | (0.271) | -0.065 | -0.200 |
| $In_3Lu$ | 4.555  | 0.238 | (0.238) | +0.029 | -0.106 |
| $Pd_3La$ | 4.235* | 0.491 | 0.991 | -0.245 | -0.130 |
| $Pd_3Ce$ | 4.135  | 0.552 | 1.052 | -0.174 | -0.059 |
| $Pd_3Lu$ | 4.033  | 0.476 | 0.976 | -0.102 | +0.010 |

Fig. 1), which can be compared with the sum $d_A + d_B$. The difference $d_A + d_B - \sqrt{2}a$ is a measure of the degree of compression of the structure caused by binding, and is equal to about 0.27 A and 0.50 A, respectively, for $In_3R$ and $Pd_3R$ phases (column 3). The degree of compression of an indium solute replacing a Pd atom in $Pd_3R$ phases is much greater (column 4). The distance between neighboring atoms on the In- or Pd-sublattice is equal to $a/\sqrt{2}$, with the difference $d_A - a/\sqrt{2}$ giving a corresponding measure of compression between A-atoms (column 5). These values are found to be mostly *negative*, indicating "open space", or less steric hindrance, between A-atom pairs. Because of size differences, there is even more open space about a Cd solute replacing a host In-atom, but less when it replaces a smaller host Pd-atom (compare columns 5 and 6). These trends support the idea that a daughter Cd-probe is more able to jump to an interstitial site in indide phases with light lanthanide elements R= La, Ce, etc.

## Relaxation Mechanisms

Most measurements on indides contributing to Figs. 4 and 5 were made in the slow-fluctuation regime, illustrated in Fig. 2 for spectra measured on In-rich $In_3La$ at 156, 261 and 340°C, with the perturbation function $G_2(t)$ is given in terms of the static perturbation function $G_2^{static}(t)$ observed at low temperature by

$$G_2(t) = \exp(-\lambda t) G_2^{static}(t). \qquad (3)$$

Here, $\lambda$ is the relaxation frequency, which damps out the static perturbation function. For relaxation arising from the model of Cd probe atoms jumping on the In-sublattice due to rapid passage of vacancies, with the EFG reorienting by 90° in each jump, the relaxation frequency is equal in very good approximation to the mean jump-frequency, or inverse of the mean residence time of the probe atom on a site [12]. Fig. 3 strictly shows relaxation frequencies $\lambda$ obtained from fits to Eq. 3 (or more accurate numerical simulations [2]), and only indirectly shows jump frequencies. But $\lambda$ must be reinterpreted for other relaxation mechanisms.

Increasing relaxation-frequencies along the indide series Nd, Pr, Ce, La shown in Fig. 4 or 5 are correlated with expanding lattice parameters and, as a consequence, greater volumes for interstitial sites. As shown elsewhere via DFT calculations for end-member $In_3La$ and $In_3Lu$ phases, Cd impurities preferentially occupy In sites over rare-earth sites [9]. But the enthalpy of a Cd impurity at an interstitial lattice location may be still lower than at either substitutional site in the "expanded" phase $In_3La$. Following decay of $^{111}In$, it is proposed that $^{111}Cd$-solutes in $In_3La$ and similar phases jump to a nearby interstitial site, most likely at or close to the octahedral interstitial site at center of the unit cell pictured in Fig. 1.

The initial jump to an interstitial site may be followed by one of several possible scenarios. Following the initial (nonstationary) jump to an interstitial location, the probe may make subsequent jumps. Fig. 6 illustrates several possible scenarios. Fig. 6a shows the Cd probe on the In-site immediately after transmutation. In Fig. 6b, the probe atom is shown displaced close to one of its two neighboring octahedral interstitial sites, where it is surrounded by five In-atoms and the vacated In-site. By symmetry, axes of the EFG at the original In-site and interstitial site are collinear, so there is no reorientation of the EFG in the initial jump although there may be a change in magnitude and/or sign.

If the migration barrier between the In-site and interstitial site is small, then the Cd-atom may rattle back-and-forth in caged motion between the original, central site and the pair of interstitial sites on either side (Fig. 6c). Finally, if a neighboring In-atom can jump into the vacated In-site (Fig. 6d), the EFG will reorient by 90 degrees. Indeed, the combination of motions shown in Figs. 6c and 6d could lead to correlated long-range migration of the Cd-impurity and its associated vacant site. These motions can lead to nuclear relaxation such as that observed.

The magnitude of observed relaxation for jumping or diffusing atoms is most likely dominated by the change in the EFG tensor in the first jump. Typical values of static quadrupole interaction frequencies of $^{111}Cd$ probes at non-cubic lattice locations in solids are of order 20-200 Mrad/s (the frequency is ~60 Mrad/s for $^{111}Cd$ on the In-sublattice in $In_3La$). Given uncertainties in changes in orientation and magnitude of EFG's in jumps, it will be assumed that the jump-frequency (inverse of the mean residence time) deduced using a generic model is uncertain by a factor of three. From Fig. 3, the fitted jump-frequency is 2 GHz at 1000 K (note a factor of two correction in the erratum in [2]), suggesting a corresponding residence time in the range 0.2 to 1.5 ns. The residence time can be that short because, following transmutation, the initial jump is not mediated by a diffusing point defect.

The lattice location of Cd appears highly unstable against transfer to an interstitial lattice location in $In_3La$. Based on a frequency of 800 Hz extrapolated to room temperature for the In-rich $In_3La$ sample in Fig. 3 [2], the mean residence time of $^{111}Cd$ on the In-sublattice in $In_3La$ is estimated to be in the range 0.4 to 4.0 ms at room temperature. For transfer to an adjacent octahedral interstitial site, there is no change in the orientation of the EFG although it might change sign. Calculations via a density function theory method such as Wien2k [13] of the stability of a Cd-interstitial adjacent to an In-vacancy and of the associated EFG at the Cd-nucleus would help elucidate the present hypothesis that the Cd-daughter transfers to an interstitial location following electron-capture decay of $^{111}In$.

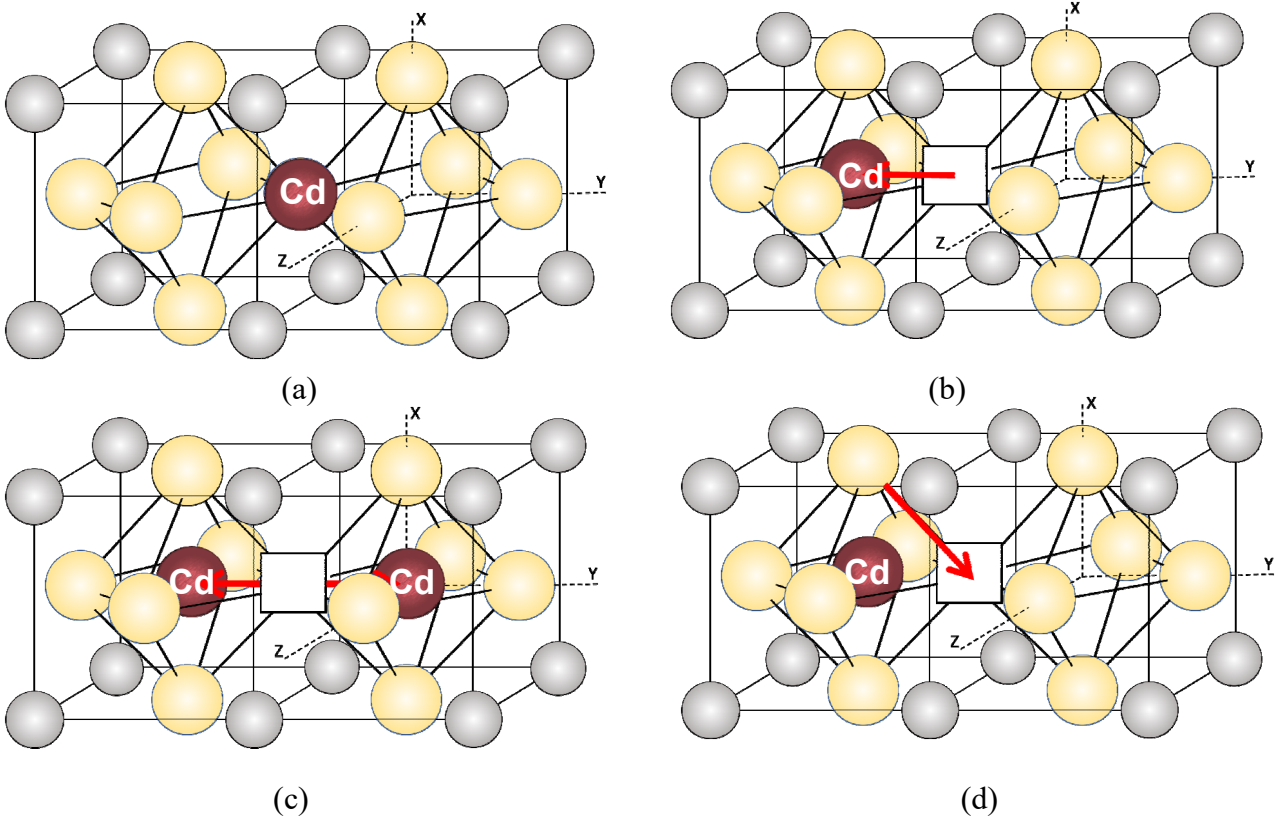

Fig. 6. Relaxation mechanisms for Cd-impurities in In$_3$La. (a) Cd-daughter on In-sublattice immediately following transmutation. (b) Cd-daughter jumps from lattice site to interstitial lattice location. (c) Cd-atom may rattle back and forth between a pair of interstitial sites. (d) In-atom may jump into site vacated by Cd-atom, leading to reorientation of the EFG at the Cd-nucleus by 90 degrees.

A final, less likely, mechanism involving neither defect-mediation nor interstitials is ring-diffusion of three- or four-atom units, with a group of atoms rotating together in ratcheted motion about an axis, e.g., a triangular group of the Cd-solute and two nearby In-atoms, or a square group of the solute and three In-atoms, all connected by near-neighbor bonds. Ring-diffusion should be more favorable for phases such as In$_3$La that have larger lattice parameters. As described above, large R-atoms open up the spacing of atoms in the unit cell, enhancing the possibility for ring-diffusion by reducing migration enthalpy barriers.

**Composition Dependence of the Jump-Rate in the Indides**

Why should the jump-rate for Cd-solutes in In$_3$La be greater for In-rich samples than for In-poor samples, as observed in Figs. 3 and 4? This is attributed to the greater chemical potential of In atoms in In-rich samples, and gives support to the interstitial transfer hypothesis.
A greater chemical potential of In increases the mole fraction of In and creates a "particle pressure" that drives solute atoms off the In-sublattice. Assume that each site on the In-sublattice is either vacant or occupied by a single In- or Cd-atom and that sites are noninteracting. The partition sum for a site can then be written in grand canonical formalism as

$$Z = 1 + \exp\{-(\varepsilon_{In} - \mu_{In})/k_B T\} + \exp\{-(\varepsilon_{Cd} - \mu_{Cd})/k_B T\}, \tag{4}$$

in which $\varepsilon_{In,Cd}$ are energies of an In or Cd atom that occupies a site and $\mu_{In,Cd}$ are the chemical potentials of In or Cd. From eq. 4, the probability that a single site is occupied by a Cd atom is

$$P_{Cd} = \frac{\exp\{-(\varepsilon_{Cd} - \mu_{Cd})/k_B T\}}{1 + \exp\{-(\varepsilon_{In} - \mu_{In})/k_B T\} + \exp\{-(\varepsilon_{Cd} - \mu_{Cd})/k_B T\}}. \qquad (5)$$

Since the mole-fraction of Cd solutes is vanishingly small and essentially invariable, the Gibbs factor in the numerator of eq. 5 has essentially a very small constant value, designated below as *B*. From eq. 4, the probability that a site is occupied by an In atom is then

$$P_{In} \cong \frac{\exp\{-(\varepsilon_{In} - \mu_{In})/k_B T\}}{1 + \exp\{-(\varepsilon_{In} - \mu_{In})/k_B T\}} = \frac{1}{\exp\{(\varepsilon_{In} - \mu_{In})/k_B T\} + 1}, \qquad (6)$$

and from eqs. 5 and 6, the ratio of probabilities is found to be equal to

$$\frac{P_{Cd}}{P_{In}} \cong \frac{B}{\exp\{-(\varepsilon_{In} - \mu_{In})/k_B T\}}. \qquad (7)$$

The ratio decreases as $\mu_{In}$ becomes more positive; that is, as the mole fraction of In increases. This argument shows why the equilibrium mole-fraction of solutes on the In-sublattice should decreases with increasing $\mu_{In}$. Chemical equilibration drives a kinetics that forces solutes away to other sublattices.

**Summary and Conclusions**

Earlier measurements along the series of In$_3$R phases suggested that there were two competing mechanisms for diffusion of $^{111}$Cd probes on the In-sublattice, dominant at opposing ends of the series [5]. More recent measurements on Pd$_3$R phases demonstrated a change in site-preference of In-solutes along the series and suggested from observed nuclear relaxation that there is also a change in site-preference of daughter Cd solutes. A similar decrease in site-preference of Cd-solutes along the In$_3$R series close to R= La is posited. The very large magnitude of relaxation in In$_3$La and similar phases appears to rule out a vacancy-mediated diffusion mechanism. The most likely mechanism has Cd-solutes created on In-sites in indides (or Pd-sites in palladides) with R= La, Ce, Pr, Nd jumping to neighboring interstitial sites [8]. Existence or not of subsequent jumps cannot be resolved from the measurements because the overall relaxation is dominated by the change in the EFG in the first jump. Given this interpretation, the fitted activation enthalpy of 0.53(1) eV for the In-rich data set shown in Fig. 3 is equal to be the migration barrier for a Cd-solute to jump between an In-site and neighboring interstitial site. The jump-frequency prefactor of 1.0 THz is consistent with the vibrational frequency of an atom in a solid. A variety of scenarios were presented that can in principle be tested by calculations of site-enthalpies and migration barriers to jumps using density function theory programs such as Wien2k.

**Acknowledgments**

Measurements on the indides were made by Aurélie Favrot, Matthew O. Zacate, Xia Jiang, John P. Bevington, Farida Selim, Li Kang, Denys Solodovnikov, Egbert Nieuwenhuis, Jipeng Wang, Megan Lockwood (Harberts), Benjamin Norman, Randal Newhouse and Justine Minish. Qiaoming Wang made the measurements on the palladides. I am indebted to them and, especially, colleague Matt Zacate for productive interactions over many years. This work was supported in part by the National Science Foundation under grant DMR 18-09531 (MMN Program) and predecessor grants.

# Appendix

The Table lists experimental values of $1/k_B T_{10}$ shown in Fig. 4, which was reproduced from Ref. 5, for In$_3$R phases having the L1$_2$ structure, in which R is a rare-earth element. These values have not previously been published. Columns 4 and 5 give values for samples having compositions at the opposing In-richer and In-poorer phase boundaries. $T_{10}$ is the temperature at which the relaxation frequency was equal to 10 MHz, obtained as shown in Fig. 2 of ref. [5]. Units of $1/k_B T_{10}$ are $eV^{-1}$, with typical uncertainties of ±0.07. Lattice parameters are from ref. [11].

| Z | R | a (nm) | $1/k_B T_{10} (eV^{-1})$ In-rich | $1/k_B T_{10} (eV^{-1})$ In-poor |
|---|---|---|---|---|
| 21 | Sc | 0.4477 | 12.21 | 12.80 |
| 39 | Y | 0.4592 | -- | 10.52 |
| 57 | La | 0.4732 | 21.52 | 14.74 |
| 58 | Ce | 0.4691 | 13.20 | 10.75 |
| 59 | Pr | 0.4671 | 10.80 | 9.73 |
| 60 | Nd | 0.4653 | 9.95 | 9.69 |
| 61 | (Pm) | -- | -- | -- |
| 62 | Sm | 0.4627 | 9.05 | 9.45 |
| 63 | (Eu) | -- | -- | -- |
| 64 | Gd | 0.4607 | 8.86 | 9.68 |
| 65 | Tb | 0.4588 | 9.08 | 9.92 |
| 66 | Dy | 0.4577 | 8.92 | 9.96 |
| 67 | Ho | 0.4573 | -- | 10.02 |
| 68 | Er | 0.4564 | 9.33 | 10.21 |
| 69 | Tm | 0.4548 | 9.93 | 10.41 |
| 70 | (Yb) | -- | -- | -- |
| 71 | Lu | 0.4544 | 10.00 | 10.65 |

# References

[1] G. Schatz and A. Weidinger, Nuclear Condensed Matter Physics, Wiley, New York, 1996.

[2] Matthew O. Zacate, Aurélie Favrot and Gary S. Collins, Atom Movement in $In_3La$ studied via Nuclear Quadrupole Relaxation, Physical Review Letters 92, 225901(1-4) (2004); *Erratum*, Physical Review Letters 93, 49903(1) (2004).

[3] Gary S. Collins, Nonstoichiometry in line compounds, Journal of Materials Science 42, 1915-1919 (2007), http://dx.doi.org/10.1007/s10853-006-0055-2).

[4] S. Maeda, K. Tanaka, and M. Koiwa, Diffusion Via Six-Jump Vacancy Cycles in the $L1_2$ Lattice, Defect Diffusion Forum 95–98, 855 (1993).

[5] Gary S. Collins, Xia Jiang, John P. Bevington, Farida Selim and Matthew O. Zacate, Change of diffusion mechanism with lattice parameter in the series of lanthanide indides having $L1_2$ structure, Physical Review Letters 102, 155901 (2009). http://link.aps.org/abstract/PRL/v102/e155901.

[6] Xia Jiang, MS thesis, Washington State University, 2008.

[7] Qiaoming Wang and Gary S. Collins, Nuclear quadrupole interactions of $^{111}$In/Cd solute atoms in a series of rare-earth palladium alloys, Hyperfine Interactions 221, 85-98 (2013). http://dx.doi.org/10.1007/s10751-012-0686-4. http://arxiv.org/abs/1209.3822.

[8] Gary S. Collins, Diffusion and equilibration of site-preferences following transmutation of tracer atoms, in Diffusion and Thermal Transport in Bulk and Nano-materials, ed. Helmut Mehrer, ISSN 2296-3642, Diffusion Foundations 19 (2019) 61-79. https://arxiv.org/abs/1805.03264

[9] M. O. Zacate, J. P. Bevington and G. S. Collins, Simulation of intrinsic defects and Cd site occupation in $LaIn_3$ and $LuIn_3$, Diffusion Foundations (this volume).

[10] Charles Kittel, Introduction to Solid State Physics, Wiley, New York 4th ed., 1971, Table 5.

[11] ASM Alloy Phase Diagram Database; https://www.asminternational.org/materials-resources/online-databases/-/journal_content/56/10192/15469013/DATABASE. The $Pd_3Ce$ lattice parameter is from Kyu S. Sim et al., J. Chem. Soc. Faraday Trans. 87 (1991) 1453-1460.

[12] A. Baudry and P. Boyer, Hyperfine Interactions 35, 803 (1987).

[13] For Wien2k, see http://susi.theochem.tuwien.ac.at/.